\documentclass[letterpaper,english,aps,prl,twocolumn,floatfix,showpacs,amsfonts,amssymb,superscriptaddress]{revtex4}

\usepackage[T1]{fontenc}
\usepackage[utf8]{inputenc}
\usepackage{graphics}
\usepackage{amssymb}
\usepackage{overpic}
\usepackage{color}
\usepackage{bm}
\usepackage{ulem}

\newcommand{\beq}{\begin{equation}}
\newcommand{\eeq}{\end{equation}}
\newcommand{\bea}{\begin{eqnarray}}
\newcommand{\eea}{\end{eqnarray}}

\begin{document}

\title{Characterizing critical 
exponents via Purcell effect}

\author{M.~B.~Silva~Neto, D.~Szilard, F.~S.~S.~Rosa, C.~Farina, and F.~A.~Pinheiro}

\affiliation{Instituto de F\' isica, Universidade Federal do Rio de Janeiro, Caixa Postal 68528, Brazil}

\begin{abstract}

We investigate the role of phase transitions into the spontaneous emission rate 
of quantum emitters embedded in a critical medium. Using a Landau-Ginzburg 
approach, we find that, in the broken symmetry phase, the 
emission rate is reduced or even 
suppressed due to the photon mass generated by the Higgs mechanism. Moreover, 
we show that the spontaneous emission presents a remarkable dependence upon the 
critical exponents associated to a given phase transition, allowing for an optical 
determination of the universality class. Our findings not only demonstrate that the Purcell effect constitutes an efficient
optical probe of distinct critical phenomena, but they also unveil that a more general connection between phase transitions and spontaneous emission exist, as previous experimental and numerical evidences suggest.
\end{abstract}

\pacs{}

\maketitle

Critical phenomena and phase transitions are amongst the most important and interdisciplinary research areas in physics.
Criticality is known to dramatically affect many structural, thermal, and electrical properties of matter~\cite{stanley}. The importance of the concept of
criticality extrapolates the domains of physics and finds applications in mathematics, biology, chemistry, and even economy and social sciences~\cite{stanley}. In addition to its phenomenological
relevance, the field of critical phenomena has always been the scenario of new and groundbreaking theoretical ideas over the years, such as renomalization group and topological phase transitions~\cite{hasan}.

In optics, critical phenomena in matter also show up in a crucial way. Important examples are the optical bistability~\cite{abraham}, the many optical manifestations of structural phase transitions in liquid crystal~\cite{singh} and, more recently, the optical analogue of the spin glass phase transition in random~\cite{ghofraniha,gomes} and homogeneous lasers~\cite{basak}. Besides, the development of structured, artificial material platforms to investigate light-matter interaction, such as photonic crystals and metamaterials, has opened new venues to investigate optical manifestations of phase transitions. For instance, manifestations of the percolation phase transition were experimentally shown to occur in the Fano lineshape that describes light reflection upon disordered photonic crytstals~\cite{pariente}.      

The high sensitivity of the spontaneous emission (SE) rate of an excited dipole 
emitter to the local environment makes the Purcell effect~\cite{purcell} especially prone to be influenced by phase transitions in 
matter. Indeed, the Purcell effect and single-molecule spectroscopy are unique 
tools to locally probe the electromagnetic environment at the 
nanoscale~\cite{moerner}, with applications in solar cells~\cite{oregan}, 
molecular imaging~\cite{moerner,vallee}, and single-photon 
sources~\cite{michler}. In addition, progress in the field of nanophotonics 
and metamaterials has allowed for unprecedented control of the 
SE rate in artificial media such as invisibility 
cloaks~\cite{wilton1}, 
graphene-based structures~\cite{wilton2}, nanoantennas~\cite{curto}, photonic 
crystals~\cite{lodahl}, and hyperbolic metamaterials~\cite{cortes}. In 
particular, the latter may undergo a topological phase transition that manifests 
itself in the Purcell factor~\cite{mirmoosa}. By inducing long-range spatial 
correlations, structural phase transitions were demonstrated to have a dramatic 
impact on the distribution of decay rates in disordered photonic 
media~\cite{juanjo1}. The decay rate of emitters embedded in a medium undergoing 
a structural phase transition induced by the temperature is also 
characteristically affected at criticality, even though other optical phenomena 
such as light scattering are insensitive to phase-switching 
behaviour~\cite{juanjo2}. Another example is the percolation transition, which 
was shown to largely enhance the decay rate of quantum emitters and crucially 
govern the decay pathways~\cite{daniela}. In addition, fluctuations of the local density of states were experimentally shown to be maximum
in thin metallic films near the percolation transition~\cite{carminati2010}.
Altogether these recent findings on 
the Purcell effect at phase transitions, of different physical origins, suggest 
that a more general and profound connection between these phenomena exists.

In order to elucidate this issue, in the present Letter we investigate the 
effects of a phase transition into the SE rate 
of emitters when embedded in a bulk critical medium. By means of a generic 
Landau-Ginzburg description we find, without specifying any particular physical 
system, that in the broken symmetry phase, the 
emission rate is reduced (or even 
suppressed) due to the photon mass generated by the Higgs mechanism. Moreover, 
we show that the spontaneous emission presents a remarkable dependence upon the 
critical exponents associated to a given phase transition, allowing for the 
determination of the universality class in the broken symmetry phase. In the symmetric phase, we show that 
enhanced critical fluctuations lead to an anomalously large enhancement of the SE at critical point.

The Purcell effect is charaterized by an enhancement of the spontaneous
emission rate, $\Gamma$, of atoms or molecules by its environment. 
For a two-level system in free space, the SE rate is
\beq
\Gamma_0=\frac{\omega_0^3 \boldsymbol{\mu}^2}{3\pi\varepsilon_0\hbar 
c^3}=\frac{\pi\omega_0 \boldsymbol{\mu}^2}{3\varepsilon_0\hbar}\rho_0(\omega_0)
\label{SER-vac}
\eeq
where $\omega_0$ is the two-level transition frequency, $\boldsymbol{\mu}$ is 
the transition dipole moment, and we identified the local density of states 
(LDOS) in vacuum $\rho_0(\omega)=\omega^2/\pi^2 c^3$. Changes in the 
electric dipole coupling and/or boundary conditions usually modify Eq.~\ref{SER-vac}, which can be more easily identified by rewriting the SE rate (divided by $2 \pi$) as
\beq
g^2 \rho_0(\omega)\equiv \frac{1}{2\hbar\varepsilon_0}
\sum_{\mu,\nu=1,2}\int\frac{d^3{\bf k}}{(2\pi)^3} 
\big\vert \hat{\epsilon}_{\bf k} \cdot \boldsymbol{\mu} \big\vert^2 \omega_{\bf 
k}^2 \, {\cal A}^{(0)}_{\mu\nu}(\omega_{\bf k},\omega),
\label{Eliashberg}
\eeq
where $g^2$ is a convenient normalization factor, $\omega_k = 
|{\bf k}| c$, $\hat{\epsilon}_k$ is the polarization versor, and ${\cal 
A}^{(0)}_{\mu\nu}(\omega_{\bf k},\omega)$ is the free-photon spectral
function 
\bea
{\cal A}^{(0)}_{\mu\nu}(\omega_{\bf k},\omega)&=&-\frac{1}{\pi}\lim_{\delta\rightarrow 0}{\cal I}m G_{\mu\nu}^{(0)}(\omega_{\bf k},\omega+i\delta) \nonumber \\
&=&\frac{\delta_{\mu\nu}}{\omega_{\bf k}}\left\{\delta(\omega-\omega_{\bf k}) - \delta(\omega+\omega_{\bf k}) \right\},
\label{SpectralFunction}
\eea
obtained from the free-photon propagator (we use the Feynman 
gauge) 
\beq
G_{\mu\nu}^{(0)}(k^2)=\frac{i\eta_{\mu\nu}}{k^2},
\eeq
where $k=(k_0,c {\bf k})$, $\eta_{\mu\nu}=diag(1,-1,-1,-1)$, and we used Lorentz invariance to simplify
the dependence of $G_{\mu\nu}^{(0)}$ to $k^{2}$.
In fact, it is easily seen that substitution of 
(\ref{SpectralFunction}) into (\ref{Eliashberg}) leads to 
(\ref{SER-vac}).

Equation (\ref{Eliashberg}) relates 
the SE rate directly to a property of its environment, in this 
case, the Green function of the quantized electromagnetic field in free space. 
For interacting fields, a natural generalization for the electromagnetic 
contribution for the SE rate then is
\beq
\Gamma\equiv 2\pi g^2 \rho(\omega_0).
\label{Gamma-from-DOS}
\eeq
where $\rho(\omega)$ is now given by an analogue of 
Eq.(\ref{Eliashberg}), but with ${\cal A}^{(0)}_{\mu\nu}$ replaced by the 
interacting electromagnetic spectral function ${\cal A}_{\mu\nu}$.

With all that in mind, we now can describe our system. The emitter is embedded in a bulk critical medium so that the Lagrangian reads
%
%This above strategy will now be used to calculate the spontaneous emission rate 
%for emitters embedded in a bulk 
%critical medium undergoing a phase transition. Let us then consider the 
%Landau-Ginzburg Lagrangian
%
\bea
{\cal L}=-\frac{1}{4}F_{\mu\nu}F^{\mu\nu}+\frac{1}{2}|D_\mu{\varphi}|^2 - a(T) \varphi^* \varphi - b (\varphi^* \varphi)^2.
\eea
Here $\varphi$ is a complex-scalar order parameter that couples to the electromagnetic
field through the covariant derivative $D_\mu=\partial_\mu - i e A_\mu$,
and the field strength tensor is $F_{\mu\nu}=\partial_\mu A_\nu-\partial_\nu A_\mu$. 
As usual, $a(T)$ is a function of $(T-T_c)$ and changes sign at the 
transition $T = T_c$; $a (T)$ and $b>0$ are the parameters that are used to 
label the different phases of the system, where $T$ is the temperature.

We proceed by: i) calculating 
$G_{\mu\nu}(k^2)$ taking into account 
the environment and/or boundary conditions; ii) extracting, from it, ${\cal 
A}_{\mu\nu}(\omega_{\bf k},\omega)$ and then $g^2\rho(\omega)$; iii) obtaining 
$\Gamma$ from Eq. (\ref{Gamma-from-DOS}). 

\begin{figure}[h]
\includegraphics[scale=0.31]{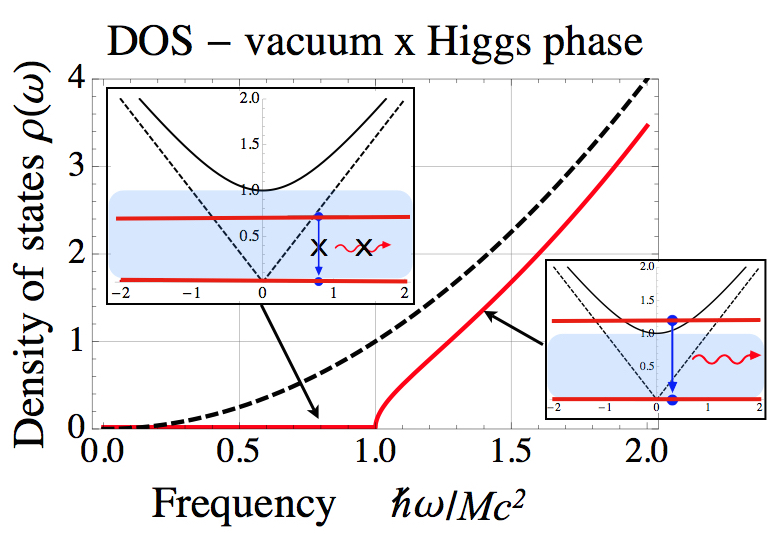}
\caption{(Main figure) photon DOS for the vacuum (black dashed curve) 
and Higgs phases (solid red curve). (Insets) for $\hbar\omega_0 < Mc^2$, inside
the shaded area, no photons are available and no emission occurs; for
$\hbar\omega_0 > Mc^2$, outside the shaded area, the DOS is finite (however low)
and emission is allowed.}
\label{Fig-DOS-Higgs}
\end{figure}

\begin{figure}[h]
\includegraphics[scale=0.6]{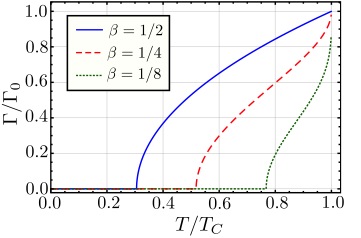}
\caption{The ratio $\Gamma/\Gamma_0$ at the Higgs phase $(T < T_c)$ for 
different typical values of the critical exponent $\beta=1/2$, $1/4$, and $1/8$.}
\label{Fig-Criticalexponents}
\end{figure}

When $T<T_c$ and $a(T)<0$, the $\varphi$ field acquires a nonzero vacuum
expectation value, $\varphi_0^2=-a/2b=v^2$,
and we need to consider perturbations around the symmetry broken vacuum, 
$\varphi (x)= e^{i \frac{\theta(x)}{v}} \left[ v+\rho(x)\right]$, where $\rho$ and $\theta$
describe longitudinal and transverse fluctuations of the order parameter $\varphi$. 
In terms of these quantities
\bea
{\cal L}&=&-\frac{1}{4}F_{\mu\nu}F^{\mu\nu}+\frac{M^2}{2}A_\mu^2 \nonumber \\
&+&\frac{1}{2}(\partial_\mu\rho)^2+\frac{m_\rho^2}{2}\rho^2+\frac{1}{2}(\partial_\mu \theta)^2+\dots
\eea
with the mass $m_\rho^2=2 |a(T)|$ being the mass of the longitudinal mode, $\rho$, 
while the transverse fluctuations, $\theta$, are massless, in accordance 
to Goldstone's theorem~\cite{peskin}. Note that a nonzero 
expectation value $v\neq 0$ provides 
the gauge field, $A_\mu$, with a mass, $M = v e$.  This is the so called Higgs mechanism~\cite{peskin}, 
in which case the massive photon propagator becomes
\beq
G_{\mu\nu}^{M}(k^2)=\frac{i\eta_{\mu\nu}}{k^2-M^2c^4/\hbar^2},
\eeq
so that the photon spectral function reads
\bea
{\cal A}^{M}_{\mu\nu}(\omega_{{\bf k},M};\omega)=
\frac{\delta_{\mu\nu}}{\omega_{{\bf k},M}}\left\{\delta(\omega-\omega_{{\bf k},M}) - \delta(\omega+\omega_{{\bf k},M}) \right\},
\label{amunu}
\eea
where the dispersion relation is ${\omega_{{\bf k},M}=\sqrt{c^2 |{\bf 
k}|^2+M^2c^4/\hbar^2}}$. The appearance of a mass term for the photons not only shifts the position
of the poles in the spectral function, but, more importantly, reduces its spectral weight.
After calculating $g^2\rho(\omega)$ from Eq. (\ref{Eliashberg}) and using
Eq. (\ref{Gamma-from-DOS}) with ${\cal A}^{M}_{\mu\nu}$ given by Eq.~(\ref{amunu}), we obtain
\beq
\Gamma=\frac{\omega_0^3\boldsymbol{\mu}^2}{3\pi\varepsilon_0\hbar c^3}\sqrt{1-\frac{M^2c^4}{\hbar^2\omega_0^2}}=
\Gamma_0\sqrt{1-\left(\frac{Mc^2}{\hbar\omega_0}\right)^2}.
\eeq
A nonzero photon mass reduces the value of the SE rate in
the Higgs phase, and 
even suppresses it, for 
$\hbar\omega_0<Mc^2$, when the energy $\hbar\omega_0$ is not large enough as 
to overcome the rest energy $Mc^2$, see Fig. \ref{Fig-DOS-Higgs}. 
%**********************************************************************************************
%**********************************************************************************************
It is important to remark that this result is valid regardless the 
specific form of the parameter $a(T)$ as long as it changes sign at 
$T_c$. Assuming a typical power law dependence ${M (t) = M_0 |t|^\beta}$ (with $t = 1 - 
T/T_c$), our findings show that, in the broken Higgs phase ($T < T_c$), the SE 
rate increases with $T$ and its behaviour crucially depends on the 
value of $\beta$ (see Fig.~\ref{Fig-Criticalexponents}). As a result, 
critical exponents $\beta$ may be easily distinguished as their effects on SE rate are 
present not only close to the transition ($T \sim T_c$), but throughout $0 < T < 
T_c$. Altogether, our results demonstrate that one can determine the universality class of an arbitrary phase transition.
This can be seen in Fig.~\ref{Fig-Criticalexponents} where the SE rate is calculated for typical values of $\beta$, such as $\beta = 1/2$ (mean field),
$\beta = 1/8$ (Ising model), $\beta = 1/4$.
%**********************************************************************************************
%**********************************************************************************************

\begin{figure}[h]
\includegraphics[scale=0.24]{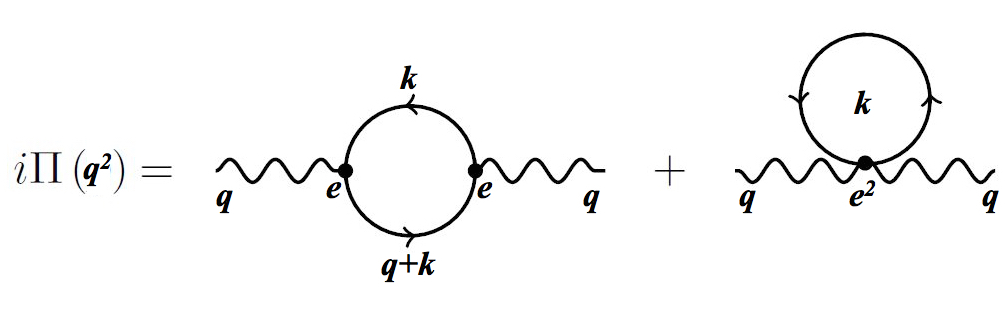}
\caption{Vacuum polarization, $\Pi(q^2)$, up to order $e^2$, from fluctuations of the complex
scalar order parameter, $\varphi^*,\varphi$.}
\label{Fig-Vacuum-Polarization}
\end{figure}

We now focus on the behaviour of the SE rate in the symmetric phase. For $T>T_c$ and $a(T)>0$, there is 
no spontaneous breakdown of the 
vacuum symmetry so $v=0$ and then 
\bea
{\cal L}=-\frac{1}{4}F_{\mu\nu}F^{\mu\nu}+\frac{1}{2}|D_\mu{\varphi}|^2 - \frac{m^2}{2} \varphi^* \varphi - b (\varphi^* \varphi)^2,
\eea
where $m=2 a(T) > 0$, for all components of $\varphi$, and we see that the photon 
is massless. Thus, in this phase, the presence of a medium surrounding the emitter does not lead to a
position shift of the pole in the photon propagator, but it rather renormalizes the 
vacuum polarization. There are two interaction vertices that 
contribute to the vacuum polarisation~\cite{greiner}
\beq
-ie(\varphi^*\partial_\mu\varphi A_\mu-(\partial_\mu\varphi)^*\varphi A_\mu),\quad\mbox{and}\quad 2i e^2 \varphi^*\varphi A^2_\mu.
\eeq
Lorentz invariance, however, constrains the vacuum polarization to
have a precise form~\cite{peskin} 
\beq
i\Pi_{\mu\nu}(q^2)=iq^2\eta_{\mu\nu}\Pi(q^2),
\eeq
where the second order contribution is shown in Fig.~\ref{Fig-Vacuum-Polarization}.

%**********************************************************************************************
%**********************************************************************************************
Going beyond the second order by summing the 
non-1PI iterations of the graphs in Fig.~\ref{Fig-Vacuum-Polarization}, leads to the 
renormalized photon propagator
\bea
G^{Z}_{\mu\nu}(q^2)&=&G^{(0)}_{\mu\nu}(q^2)+G^{(0)}_{\mu\alpha}(q^2)\Pi_{\alpha\beta}(q^2)G^{(0)}_{\beta\nu}(q^2)+\dots \nonumber\\
&=&Z(q^2) G^{(0)}_{\mu\nu}(q^2),
\eea
with
\beq
Z(q^2) = \frac{1}{1-\Pi(q^2)}.
\eeq
The function $\Pi(q^2)$ is well known and it is regular at $q=0$, thus ensuring that the photon 
remains massless. In the limit $q^2 \gg m^2$, and setting $q^2=\Lambda^2$, it is given by~\cite{peskin}
\beq
\Pi(q^2=\Lambda^2)=\frac{\alpha}{12\pi}\ln{(\Lambda^2/m^2)},
\eeq
where $\Lambda$ is a characteristic momentum scale in the problem and $\alpha=1/137$ is
the fine structure constant. 

For the case of standard quantum electrodynamics, where $m=m_e$ 
is the electron mass, one 
usually choses the {\it on shell} renormalization point, $\Lambda=m_e$, in such a way that 
$\Pi(q^2=m_e^2)=0$ and $Z(q^2=m_e^2)=1$. In solid state systems, however, where the 
mass term corresponds instead to the inverse of a correlation length, 
$m=\xi^{-1}$, the scale $\Lambda$ corresponds to a given natural cutoff 
in the problem (e.g., the Fermi momentum $k_F$, or the Debye 
momentum $q_D$, or, yet, the inverse lattice 
spacing $1/a$, among others). Here we 
choose $\Lambda=1/\xi_0$, where $\xi_0$ is some fixed length scale far away 
from the critical point, $\xi(T\gg T_c)\approx\xi_0$,
in such a way that, for $T\gg T_c$, we end up with $\Pi(q^2=\xi_0^{-2})=0$ and $Z(q^2=\xi_0^{-2})=1$.

The photon spectral function can be calculated as
\bea
{\cal A}^{Z}_{\mu\nu}(\omega_{\bf k};\omega) & = & -\frac{1}{\pi}\lim_{\delta\rightarrow 0}{\cal I}m G^{Z}_{\mu\nu}(\omega_{\bf k},\omega+i\delta) \nonumber \\
&=&Z(\xi^2){\cal A}^{(0)}_{\mu\nu}(\omega_{\bf k};\omega),
\eea
and, as a consequence, the SE rate becomes
\beq
\Gamma=Z(\xi^2/\xi_0^2)\frac{\pi\omega_0\boldsymbol{\mu}^2}{3\varepsilon_0\hbar}\rho(\omega_0)=Z(\xi^2/\xi_0^2)\Gamma_0.
\label{gammafinal}
\eeq
We clearly see that for $T\gg T_c$, when $\xi\rightarrow\xi_0$ and $Z\rightarrow 1$, we have 
$\Gamma\rightarrow\Gamma_0$. As the critical point is approached, where 
$T \rightarrow T_c$ and $\xi\gg\xi_0$, one finds $Z \gg 1$, leading to a large 
enhacement of the SE rate. The divergence in the SE rate, Eq.~(\ref{gammafinal}), can 
be understood as a result of the unscreening of the electric charge when $\xi\rightarrow\infty$. 
%{\color{red} Then, for a fixed length scale, 
%$\xi_0=1/\Lambda$, the divergence of the correlation length, as criticality
%s approached ($\xi\rightarrow\infty$), scales out the entire system, in such a 
%way that it is equivalent to looking deeper 
%inside the screening cloud, where the value of the electric charge 
%increases.{\color{blue} - ISSO ESTA UM POUCO CONFUSO, E ACHAMOS QUE NAO ESTA 
%ADICIONANDO MUITO. MELHORAR OU RETIRAR.}} This is depicted in Fig. 
%\ref{Fig-Unscreening-of-e}.  
%**********************************************************************************************
%\begin{figure}[h]
%\includegraphics[scale=0.36]{Figure-DOS-Normal.jpg}
%\caption{(Main figure) photon DOS for the vacuum (black dashed curve) 
%and symmetric phases (solid red curve). (Inset) for any $\hbar\omega_0$ the DOS
%is finite and large, as $T_c\leftarrow T$, due to critical fluctuations of the 
%soft order parameter.}
%\label{Fig-DOS-Normal}
%\end{figure}
%**********************************************************************************************
%**********************************************************************************************
%\begin{figure}[h]
%\includegraphics[scale=0.32]{Unscreening-of-e.jpg}
%\caption{The renormalized charge takes into account the screening of the bare charge,
%due to the vacuum polarisation. Divergence of the correlation length stretches out space
%exposing the bare charge and producing a divergence in the value of $e$.}
%\label{Fig-Unscreening-of-e}
%\end{figure}
%**********************************************************************************************

\begin{figure}[h]
\includegraphics[scale=0.6]{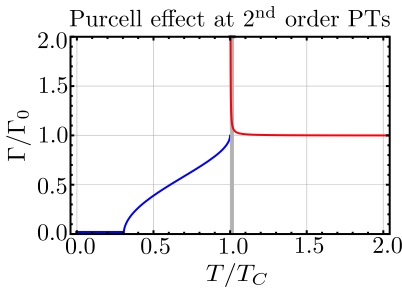}
\caption{Purcell effect close to a 2nd order PT at $T_c$. The ratio $\Gamma/\Gamma_0$ is shown 
for the two phases: i) Higgs phase (in blue) both for $\hbar\omega_0<M(T)c^2$, where $\Gamma=0$, 
or $\hbar\omega_0>M(T)c^2$, where $\Gamma<\Gamma_0$; ii) symmetric phase (in red), where 
$\Gamma$ diverges as $\xi\rightarrow\infty$, close to the stable IR fixed point.}
\label{Fig-Purcell-PT}
\end{figure}

The correlation length $\xi$ may approach criticality through a power law 
 $\xi(t)=A |t|^{-\nu}+\xi_0$, or through another function such as 
$\xi(t) = \xi_0\exp{(\delta/|t|)}$.
Due to the logarithmic dependence in $Z$, any choice of power law
for $\xi$ in the symmetric phase does not lead to significant 
effects. In order to illustrate that, in Fig. \ref{Fig-Purcell-PT}
we plot the normalized SE rate with $M(t)= M_0 |t|^{1/2}$ in the Higgs 
phase, and $\xi(t) = \xi_0\exp{(10/|t|)}$ for the symmetric phase (recalling 
that $t=1-T/T_c$). It is clear that, coming from the symmetric phase, the 
enhancement is huge close to $T_c$.

%**********************************************************************************************
%**********************************************************************************************

In conclusion, we have investigated, using a Landau-Ginzburg 
approach, the effects of phase transitions in the SE rate of quantum emitters
embedded in a critical medium. In the broken symmetry phase, we
demonstrate that the SE rate is reduced and even suppressed due to the photon mass generated by the Higgs mechanism.
In this same phase we show that the SE crucially depends on critical exponents associated to a given phase transition, without specifying {\it a priori} any particular physical system. This fact allows
one to determine the universality class of phase transitions by optical means. Our results show that a general connection between critical phenomena and the Purcell effect exists,
as previous experimental and numerical evidences, involving different phase transitions, suggest~\cite{juanjo1,juanjo2,daniela,carminati2010}. Hence our analysis provide a qualitative theoretical basis for these findings, revealing they are actually deeply related. Altogether, our work not only unveils the general connection between critical phenomena and the Purcell effect, but it also demonstrates that the latter may be exploited as an optical probe of phase transitions and their universality classes.   

\acknowledgements 
The authors are grateful to N. de Sousa, J. J. Saenz, C. Lopez, L. Moriconi, R. de Melo e Souza for 
valuable suggestions. D.S., F.S.S.R., and C.F. 
acknowledge CAPES, CNPq, and FAPERJ for partially financing this research. F.A.P. 
acknowledges the financial support of the Royal Society (U.K.) through a Newton 
Advanced Fellowship (Ref. No. NA150208), FAPERJ (APQ1-210.611/2016) and CNPq 
(Grant No. 303286/2013-0).

\end{document}